\documentclass[preprint2]{aastex63}
\usepackage{hyperref}
\usepackage{soul}
\usepackage{lineno}

\def \sgr {\mbox{SGR~1935$+$2154}}
\def \frb  {\mbox{FRB~20200120E}}

\def\ltsima{$\; \buildrel < \over \sim \;$}
\def\lsim{\lower.5ex\hbox{\ltsima}}
\def\gtsima{$\; \buildrel > \over \sim \;$}
\def\gsim{\lower.5ex\hbox{\gtsima}}

\received{2021 September 17 }
\revised{2021 October 12}
\accepted{2021 October 12}
\submitjournal{ApJ Letters}

\shorttitle{}
\shortauthors{Mereghetti et al.}

\begin{document}

\title{INTEGRAL limits on past high-energy activity from     \frb\ in M81}

\author[0000-0003-3259-7801]{S.~Mereghetti} 
\affiliation{INAF -- Istituto di Astrofisica Spaziale e Fisica Cosmica, Via A. Corti 12, I-20133 Milano, Italy}

\author[0000-0002-5711-9278]{M.~Topinka} 
\affiliation{INAF -- Istituto di Astrofisica Spaziale e Fisica Cosmica, Via A. Corti 12, I-20133 Milano, Italy}

\author[0000-0001-6641-5450]{M.~Rigoselli} 
\affiliation{INAF -- Istituto di Astrofisica Spaziale e Fisica Cosmica, Via A. Corti 12, I-20133 Milano, Italy}
 
\author[0000-0001-9494-0981]{D.~G\"otz} 
\affiliation{AIM-CEA/DRF/Irfu/D\'epartement d’Astrophysique, CNRS, Universit\'e Paris-Saclay, Universit\'e de Paris, \\
\hspace{0.05cm}Orme des Merisiers, F-91191 Gif-sur-Yvette, France}

\begin{abstract}
The  repeating fast radio burst \frb\ is located in a globular cluster belonging to the  nearby M81 galaxy. Its small distance  (3.6 Mpc) and accurate localization make it an interesting target to  search for bursting activity at high energies. From   November 2003  to  September 2021, the INTEGRAL satellite has obtained an exposure time of  18 Ms on the M81 sky region.  We used these data to search for hard X-ray bursts from \frb\ using   the  IBIS/ISGRI instrument, without finding any significant  candidate, down to  an  average  fluence limit of $\sim10^{-8}$ erg cm$^{-2}$  (20-200 keV). The corresponding limit on the isotropic luminosity for a burst of duration $\Delta t$ is $\sim10^{45} \left ( \frac{10~ms}{\Delta t} \right )$ erg s$^{-1}$, the deepest limit obtained for an extragalactic FRB in the hard X-ray range. This rules out the emission of powerful flares  at a rate higher than 0.1 yr$^{-1}$ that could be expected in models invoking young hyper-active magnetars. 
\end{abstract}



\section{Introduction}
\label{sec:intro}

The discovery of an extremely bright and short radio burst from the Galactic soft gamma-ray repeater \sgr\ on 28 April   2020 \citep{CHI20b,boc20},   provided strong observational support to the connection between  fast radio bursts (FRBs)  and magnetars. FRBs are short ($\sim$ms) and bright pulses of coherent radio emission with high dispersion measure, implying an extragalactic origin (see \citealt{pet19,cor19} for reviews), while magnetars are isolated neutron stars powered by magnetic energy (see \citealt{mer15,tur15,kas17} for reviews).  A connection with magnetars had been postulated in several of the FRB models, involving emission either in the star magnetosphere  or in relativistically ejected material (see, e.g.,  \citealt{zha20} and references therein). 

The bright FRB-like radio burst of   28 April  2020  from \sgr\   was accompanied by the simultaneous   emission of hard X-rays with properties similar to those of the  short  bursts  (duration $<$1 s, peak luminosity $\sim10^{39-41}$ erg s$^{-1}$) typical  of  \sgr\  and other Galactic magnetars, except for a  harder spectrum \citep{mer20,rid21,li21}.  The ratio between the radio and X-ray fluences of the 28 April 2020 burst was $\eta\sim3\times10^{-5}$.   Fainter radio bursts  subsequently observed from \sgr\ were not seen at high energies \citep{kir21n},   consistent with similar values of $\eta$.  On the other hand,  the lack of a radio burst associated with the 2004 December 27 giant flare of  SGR 1806--20  \citep{pal05} implies  $\eta<10^{-10}$ \citep{ten16},  indicating that the ratio of radio to high-energy fluence in magnetar bursts  can span a wide range of values.

Searches for X-ray counterparts of FRBs, already carried out in the past without success \citep{sch17,cun19,gui19,mar19}, received renewed interest after the \sgr\ results.  However, the extragalactic    distances of   FRBs result in much lower fluxes compared to those of galactic magnetars (\sgr\ has a  distance of   4.4$^{+2.8}_{-1.3}$ kpc,  \citealt{mer20}). Up to now, only upper limits have been reported for their high-energy emission \citep{sch20,gui20,pil20,ver21}.

Recently, the repeating  \frb\ has been discovered in the outskirts of the  spiral galaxy  M81 \citep{bha21}.   Its association with this nearby galaxy,  already suggested by the  FRB dispersion measure of only 88 pc cm$^{-3}$, has been confirmed  by  a sub-arcsecond localization that shows positional coincidence with a globular cluster belonging to M81 \citep{kir21}. With a distance of only 3.6 Mpc, \frb\ is by far the closest extragalactic FRB. Therefore, also considering its repeating nature, it is a very interesting target for multiwavelength observations.  These observations allow to sample luminosities $\sim$2000 times below the limits obtained for FRB 20180916B, which is the next closest repeating FRB with an  identified host galaxy (d$\sim$150 Mpc, \citealt{mar20}).   We note that another interesting target in this respect is  FRB 20181030A, if its recently suggested association with NGC 3252 at $\sim$20 Mpc \citep{bha21b} is confirmed.  
A search for persistent X-ray sources at the position of \frb\ was done with Chandra by \citet{kir21}, who obtained a luminosity upper limit of 2$\times$10$^{37}$ erg s$^{-1}$ (0.5--10 keV). They also noticed the lack of  $\gamma$-ray  sources in the  Fermi/LAT catalogues at the M81 position.  

Here we report on  a search for hard X-ray  bursts from \frb\ carried out using the archival data of  the INTEGRAL satellite, which collected an exposure time of    18 millions of seconds on M81 from   November 2003 to September 2021.

\begin{table*}[h!]

\caption{Observations log
\label{tab-obslog}
}
 \begin{center}
 \begin{tabular}{ccccrr}
 \hline
\smallskip
     Revolutions & Start Date & End Date &   Off-axis angles            & Net exposure   & Net exposure\\ 
                       &                  &                &                                     & Fully coded FoV    & Total     \\
                     &                  &                &   (degrees)                     &  (ks) & (ks)     \\
\hline
132-133 & 2003-11-12 & 2003-11-17 &  9.3--15.6 &      0 &    197 \\
179-180 & 2004-04-02 & 2004-04-06 &  9.0--16.3 &      0 &    141 \\
221-224 & 2004-08-04 & 2004-08-14 & 11.8--13.8 &      0 &     92 \\
250 & 2004-10-30 & 2004-10-31 &  7.9--8.0 &      0 &      4 \\
669 & 2008-04-07 & 2008-04-07 & 16.7--16.7 &      0 &      1 \\
856-872 & 2009-10-17 & 2009-12-06 &  0.3--8.1 &   1077 &   1626 \\
930-933 & 2010-05-28 & 2010-06-05 &  0.5--9.8 &    154 &    241 \\
960-962 & 2010-08-24 & 2010-08-31 & 13.3--15.1 &      0 &     10 \\
971-977 & 2010-09-26 & 2010-10-16 &  0.9--15.0 &     73 &    108 \\
1029-1051 & 2011-03-18 & 2011-05-24 &  0.4--8.3 &    602 &    883 \\
1092-1096 & 2011-09-24 & 2011-10-06 &  1.6--7.9 &     39 &     61 \\
1111-1115 & 2011-11-20 & 2011-12-02 &  0.6--8.0 &    316 &    495 \\
1156 & 2012-04-03 & 2012-04-03 &  8.0--8.1 &      0 &      5 \\
1173 & 2012-05-22 & 2012-05-22 & 14.4--14.7 &      0 &      2 \\
1216-1254 & 2012-09-30 & 2013-01-20 &  0.3--8.1 &   1107 &   1681 \\
1347-1364 & 2013-10-25 & 2013-12-16 &  0.3--20.5 &   1179 &   2162 \\
1380-1407 & 2014-01-31 & 2014-04-24 &  0.0--9.6 &   2916 &   4512 \\
1419-1432 & 2014-05-27 & 2014-07-08 &  0.1--8.2 &   1432 &   2154 \\
1524 & 2015-03-30 & 2015-03-30 &  7.9--8.0 &      0 &      8 \\
1558-1564 & 2015-06-30 & 2015-07-16 & 10.5--10.7 &      0 &      3 \\
1652 & 2016-03-05 & 2016-03-05 &  8.1--8.2 &      0 &      6 \\
1671-1681 & 2016-04-25 & 2016-05-22 & 11.5--16.0 &      0 &    282 \\
1852 & 2017-08-20 & 2017-08-20 &  2.9--3.9 &      1 &      1 \\
1873-1874 & 2017-10-15 & 2017-10-18 &  3.9--4.0 &      3 &      3 \\
2117-2119 & 2019-07-26 & 2019-07-29 & 14.3--14.9 &      0 &     11 \\
2151-2167 & 2019-10-24 & 2019-12-04 &  0.5--13.5 &    808 &   1204 \\
2227-2228 & 2020-05-13 & 2020-05-16 & 1.2--8.0 &       102 &   151 \\
2278-2307 & 2020-09-24 & 2020-12-11 &  0.7--8.1 & 502 & 727 \\
2334-2372 & 2021-02-21 & 2021-06-03 & 0.4--17.7 & 785& 1149 \\
2395-2410 & 2021-08-02 & 2021-09-11 & 0.3--6.9 & 100 & 150  \\

\hline
 TOTAL     &                          &                   &               &  11196            &   18067          \\
\hline
\smallskip
 \end{tabular}
\end{center}
\end{table*}


\begin{figure*}[ht!]
\includegraphics[width=\textwidth]{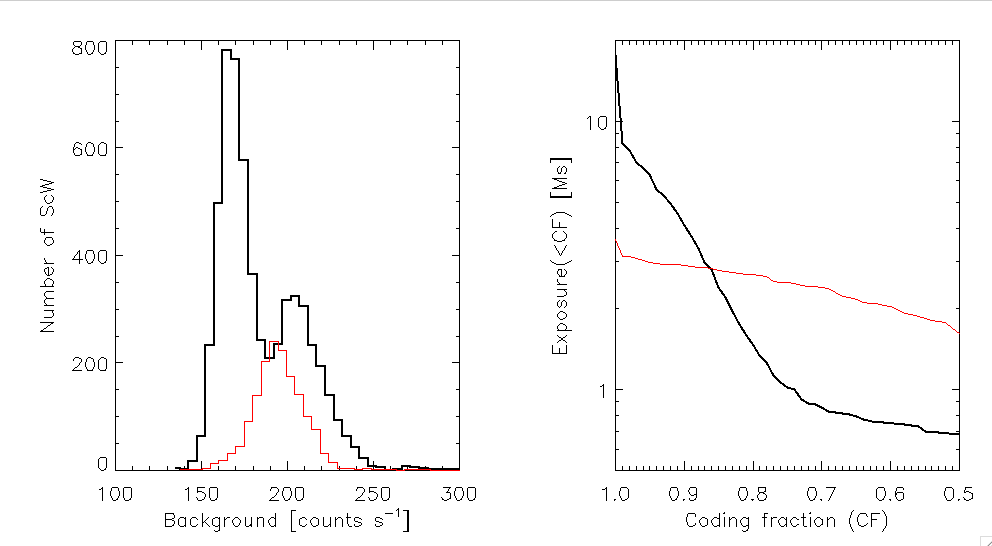}
\caption{Left: distribution of background count rates  (20-200 keV) of the ScW used in the search for bursts from \frb\ (black) and in the observations of SGR 1806--20 used for comparison (red).   The count rates have been corrected for the coding fraction (i.e. the fraction of detector area over which the photons coming from the source direction are  modulated by the coded mask aperture pattern). 
Right:   distribution of net exposure time  as a function of coding fraction  for  \frb\ (black) and for SGR 1806--20 (red). 
\label{fig-bkg-cod} }
\end{figure*}

\begin{figure}[ht!]
\includegraphics[width=\columnwidth]{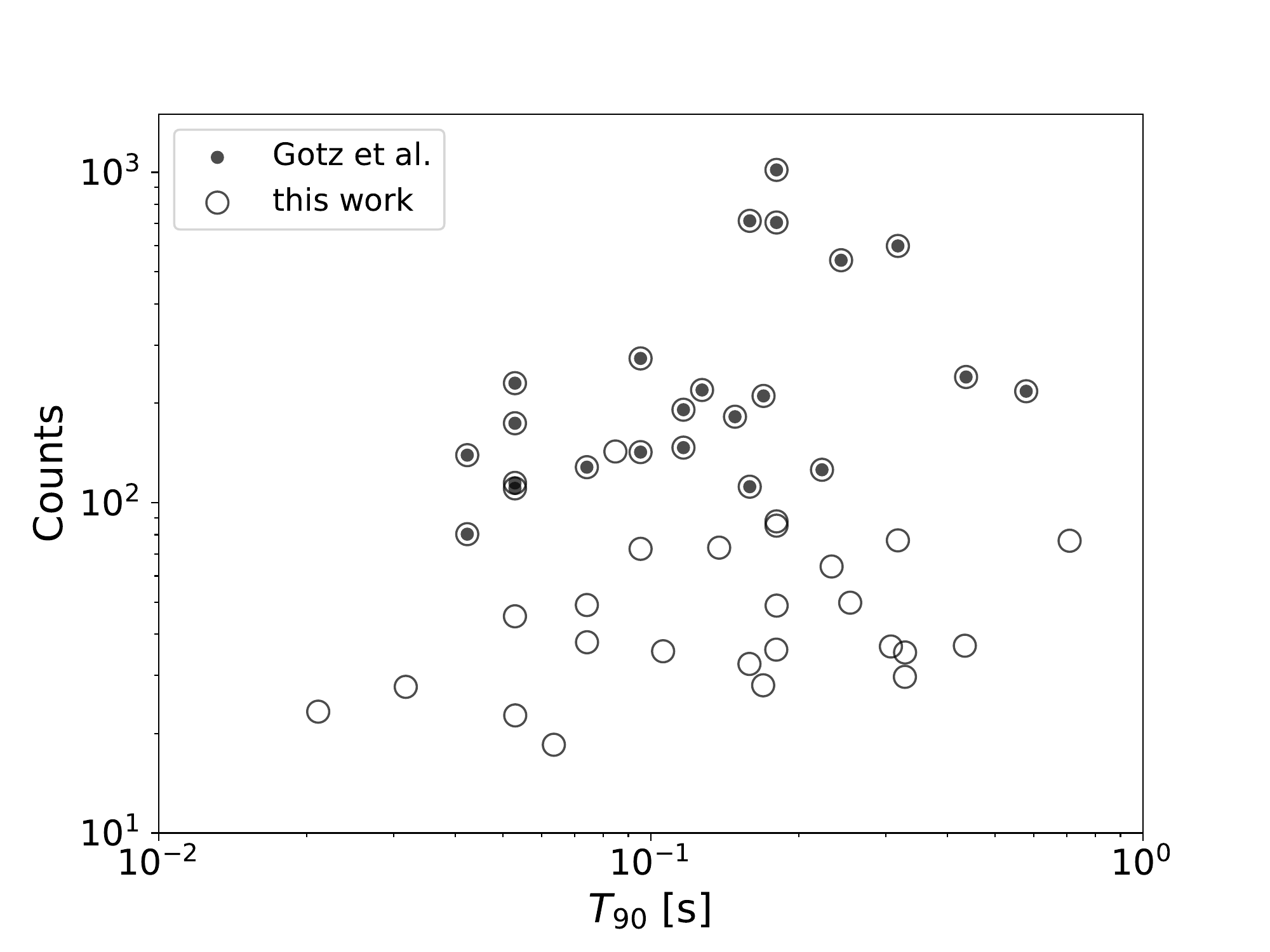}
\caption{The circles show the number of counts and duration of the bursts from SGR 1806--20  detected during revolutions 108-122 (from September 1 to October 15 2003). The dots  mark the bursts also detected by \citet{goe04b}. 
\label{fig-cts-t90} }
\end{figure}

\section{Data analysis and results} 
\label{sec:data analysis}

Our results are based on data collected by the  Imager on-board INTEGRAL (IBIS, \citealt{ube03}). IBIS is a coded mask instrument providing images  with  angular resolution of $\sim$12 arcmin over a total field of view of  $29^{\circ}\times29^{\circ}$.  In particular, we used the data obtained with ISGRI (INTEGRAL Soft Gamma Ray Imager, \citealt{leb03}), the lower energy detector of IBIS, which operates in the nominal 15-1000 keV  range and provides  photon-by-photon events tagged with a  time resolution of 61 $\mu$s.  The ISGRI detection plane consists of 128$\times$128 pixels, grouped in eight modules, giving a total sensitive area of 2600 cm$^2$ on-axis. For sources located in the  central $9^{\circ}\times9^{\circ}$ of the field of view (FoV),  the whole detector area receives flux modulated by the coded mask. In this region (the fully coded FoV)  the sensitivity is  highest and nearly uniform. For sources at larger off-axis angles (the partially coded FoV) only a fraction of the detection plane receives  the flux modulated by the coded mask, and thus the sensitivity gradually decreases.

The position of \frb\ has been repeatedly observed by INTEGRAL,  with long  campaigns dedicated to M81 \citep{meremi16}, as well as serendipitously,  during observations of other nearby targets. Due to the dithering observing mode used by INTEGRAL, the  data are split into many  short pointings with typical duration between 30 and 60 minutes each, called science windows (ScWs).  We selected for  our analysis all the ScWs that contain the position of \frb ,  including those in which it was located at  large off-axis angles.  The data were retrieved from the local archive of public INTEGRAL data maintained at IASF-Milano \citep{pai13}.
Our search consisted of two main steps: the  first one, based only on the timing analysis of the ISGRI light curves, aimed at finding burst candidates; the second step was an imaging analysis of the    candidates, in order to understand their nature and possibly associate them with \frb\ or other sources in the instrument FoV.

To increase the signal to noise ratio, all the light curves were extracted using only the ISGRI pixels with a source illumination fraction greater than 0.5. 
For most of the time, the instrumental background induced by cosmic rays, as well as  that caused by bright X-ray sources inside (or close to) the  FoV, vary  on timescales longer than the ScW duration.
ScWs with particularly high and variable background typically occur when INTEGRAL is close to the Earth radiation belts, at the beginning or at the end of the satellite revolutions\footnote{INTEGRAL is on a highly elliptical orbit with period of 3 days until  2015 January and 2.7 days afterward.}, or during periods of intense solar activity.  
Such ScWs  were excluded from our search, resulting in the net exposure times reported in Table~\ref{tab-obslog}. 
 
For each ScW, we determined the background by fitting a constant to the light curve  of the whole detector binned  at one second, after  correcting for time intervals in which one or more ISGRI modules were not operating\footnote{Modules with too many noisy pixels are temporarily switched off  by the on-board software} and iteratively removing the bins   more than three standard deviations above the average.  Such   ``3-$\sigma$ clipping'' is done to have a better estimate of the background level during the ``quiet'' part of the ScW. Inclusion of bright bins (often caused by  bursts from  sources outside the field of view) would lead to an increased threshold level, thus reducing the sensitivity. Note that the time intervals with high count rate are removed only in the background computation, while they are kept for the following burst search.   

We then searched for excesses above the expected background counts using  sliding time windows with eight durations logarithmically spaced  between 0.01 and 1.28 s.  These are the typical durations of bursts from magnetars. Longer events would trigger on several adjacent bins, unless they have a very slow rise time (which would be atypical for a magnetar burst). Events shorter than 0.01 s and with a sufficiently high number of counts to be detected also in the images ($\sim$30--100, depending on the position in the FoV)  trigger also on the longer timescales.   The thresholds were set to values corresponding to  an expectation of $\sim$0.001 false positives per ScW and timescale.  The search was done in two energy ranges: 20-200 keV   and 20-100 keV.  Excesses found in adjacent time bins and/or in overlapping timescales were grouped and considered as a single  burst candidate for further imaging analysis.

The candidates found in the light curves were   examined with the imaging software developed for the INTEGRAL Burst Alert System (IBAS, \citealt{mer03}) as well as with  a maximum likelihood method which exploits the  knowledge of the source position in the field of view. Briefly, this consists in finding the source and background fluxes which maximize the probability of obtaing the observed distribution of detector counts as a function of the pixel illumination fraction.  The time integration for these analysis was optimized for each candidate by selecting only the time intervals with a number of counts above three standard deviations from the average in the light curves. 
None of the  candidates could be confirmed as a burst positionally coincident with \frb . The only candidate producing a source significantly detected
in the images was  GRB 121212A \citep{gcn14064}. This burst,  at  angular distance of 12$^{\circ}$ from \frb\   and at   off-axis angle of  6$^{\circ}$, had previously been detected   in real time as a sub-threshold IBAS trigger \citep{hig17}.

Although the  faintest  candidates had a number of counts below that required for imaging analysis,  more than 200 of them were bright enough and should have produced a clearly detectable source,  if originating from directions within the FoV.  The most likely explanations for these events is that they were due to either background fluctuations or   bursts from sources at large off-axis angles,  outside the imaging FoV. In fact, the passive shielding that connects ISGRI to the coded mask becomes progressively transparent with increasing photon energy. In a few cases we  could associate the events with   GRBs seen by other satellites, based on their temporal coincidence.  Two such examples are  GRB 121118B  \citep{gcn13977} and GRB 140306A \citep{gcn15938}, which occurred at  off-axis angles of 73$^{\circ}$  and 50$^{\circ}$, respectively. 

Several factors concur to determine the sensitivity of our search: the most relevant one  is  the FRB position in the FoV, which determines the  detecor area collecting coded source photons.  Other factors are the background  rate and   the secular increase in the low-energy threshold caused by the detector ageing. Finally,  also  the   spectral shape,  duration and time profile of the burst have an effect on the sensitivity, but these properties do not vary much for the typical magnetar bursts.  The distribution of the background count rates in the ScWs used in our search  (left panel of Fig.~\ref{fig-bkg-cod})  has two peaks, at $\sim$170 and  $\sim$200  counts s$^{-1}$. On long timescales, the background variations follow the 11 yr cycle of solar activity. The background is  higher at the minimum of solar activity. The bimodal shape of the distribution shown in the figure reflects how the ScWs were distributed in time.   As it can be seen in Table~\ref{tab-obslog} and Fig.~\ref{fig-bkg-cod} (right panel), \frb\ was observed at small off-axis angles for most of the time: only for $\sim$1 (0.7) Ms  it  was  at positions with   coding fraction below 0.75 (0.5),  where the sensitivity is  $<$85\%  (75\%) of the optimal on-axis value.   
It is thus clear from the distributions of  Fig.~\ref{fig-bkg-cod} that the ScWs used in our search had different sensitivities.

 To test our burst search procedure and to  estimate its  sensitivity, we applied it to the    ISGRI data  obtained      in 2003-2004 for the magnetar SGR 1806$-$20, during a period of bursting activity \citep{goe04,goe06}.
Compared to the M81 data, these observations   include a higher fraction of ScWs with the source at large off-axis angles and,   due to the presence of the Galactic Ridge diffuse emission, their background rate is generally higher than that of the M81 data taken at the same phase of the 11 yr solar cycle.  As shown by the red histogram  in Fig.~\ref{fig-bkg-cod}, 
most of the M81 ScW had a background rate similar or lower than that of the  SGR 1806$-$20 data used for comparison (red histogram  in Fig.~\ref{fig-bkg-cod}).  Therefore,  using   these  SGR 1806$-$20 data we obtain a conservative estimate of the average sensitivity reached in our search.

 We detected from SGR 1806--20 almost twice as many  bursts that had been   reported by  these authors. 
As an example, we show in Fig.~\ref{fig-cts-t90} the results  for  the time period from September 1 to October 15, 2003,   that  can be directly compared to those of  \citet{goe04b}.  
Our search algorithm is able to reveal fainter bursts because the previous works were based on the triggers found during  the IBAS real time analysis.  IBAS monitors the overall count rate of the whole detector, without exploiting the {\it a priori} knowledge of the source position to optimize the light curves extraction (as instead we do in this work).    
\citet{goe06}, assuming  a thermal bremsstrahlung spectrum with temperature  kT=32 keV, measured a  15-100 keV  fluence of   $\sim10^{-8}$ erg cm$^{-2}$ for their faintest bursts.  Converting to the 20-200 keV range and considering  kT in the  20--70 keV  range  would change this value by less than 30\%.   
Therefore,  although we could reveal also fainter bursts from SGR 1806--20  and the  data on M81 had more favourable  background and coding fraction conditions, we conservatively take  $10^{-8}$ erg cm$^{-2}$ (20--200 keV) as a representative fluence upper limit valid for the bulk of the \frb\  observations reported here. \\
  
\newpage
    
\section{Discussion} 
\label{sec:discussion}

We carried out a  search for hard X-ray ($>$20 keV) bursts from the direction of \frb\ in     18 Ms  of INTEGRAL data without finding any significant candidate. This non-detection rules out  the emission of bursts with fluence above  $\sim10^{-8}$ erg cm$^{-2}$ during most of our observations.
At the distance of 3.6 Mpc, implied by the location of \frb\ in the core of  a globular cluster of  M81,  this   corresponds to a limit on the isotropic equivalent energy  of $\sim10^{43}$  erg, the   deepest limit obtained for a  FRB in the  hard X-ray range.
 The  bursts from \frb\  observed at 400-800 MHz   with CHIME  in 2020 \citep{bha21} and the one of  2021 March 2 seen at 2.25 GHz \citep{maj21}, with isotropic radio energies of  $\sim(1-2)\times10^{34}$ erg,  are the most energetic ones seen  from this repeater. Other less energetic bursts were seen at 1.4  GHz \citep{nim21}.   Unfortunatley, the time intervals of our search do not include any of these bursts.   
If   bursts of similar radio energy occurred during the INTEGRAL observations analysed here,  our limit would imply a ratio of radio to X-ray fluence $\eta \gsim 10^{-9}$.

Our  limit cannot rule out  short SGR-like   bursts from \frb . The energy emitted in typical magnetar bursts, with durations shorter than one second,  rarely exceeds  a few $10^{41}$ erg  \citep{isr08,vdh12,lin20,you20}. 
The   2020 April 28 burst from SGR 1935$+$2154 had a hard X-ray energy of $\sim3\times10^{39} \left(\frac{{\rm d}}{5~kpc}\right)^2$ erg and $\eta \sim3\times10^{-5}$ \citep{mer20,rid21}.  On the other hand, the three giant flares observed from SGR 0526--66,  SGR 1900$+$14 and SGR 1806--20  emitted   $10^{44-46}$  erg   in their short  ($<$0.4 s) initial spikes \citep{maz99,hur05}.  
Similar events at the distance of M81 would have been clearly detected in our search\footnote{\citet{bha21} noticed that the burst of 2020 July 18 was in the FoV of Swift/BAT and the candidate burst of 2020 February 6 was visible by Fermi/GBM, but no detections with these instruments were reported. The estimated limits rule out a magnetar giant flare emission associated to these two events.}. We note, however,  that  the net exposure time of about half an year obtained for \frb\ with INTEGRAL in   18  years, is   relatively small if one considers that only three giant flares  have been detected from the Galactic magnetars observed (although with discontinuous monitoring) for $\sim$40 years.

The lack of bursts from \frb\ in the INTEGRAL data is  constraining in the context of models involving  young and very active magnetars. Several FRB models
 invoke the presence of magnetars with ages below a few tens of years,  much younger  than those observed in our Galaxy \citep{met17,bel20,lu20,dal21}. The latter have estimated ages of  several tens of kyr or longer,  although some possibly younger magnetars have been recently found \citep{esp20}.   Young and fastly spinning magnetars are believed to  emit bursts and flares   more frequently and with higher energy than the older known magnetars. In fact, the activity level, as measured by the frequency and energetics of bursts and outbursts,  declines with time, as the very large initial rotational energy decreases and  the magnetic field stored in the star dissipates \citep{per11,deh20}.   The probability of observing zero bursts in 0.5 yr of data is   $e^{-R/2}$, where $R$ is the burst rate per year.  Our null results  imply  $R<0.1$ yr$^{-1}$   (at 95\% c.l.)  for bursts with luminosity above  $\sim10^{45} \left ( \frac{10~ms}{\Delta t} \right )$ erg s$^{-1}$, where $\Delta$t is the burst duration. 

We finally note that  the  location in  a globular cluster with estimated age of $\sim$10 Gyr \citep{kir21} makes it   unlikely   that \frb\ was  born in the collapse of a massive star, the favorite  channel proposed for the formation of magnetars. Possible alternative origins are  the merging of   two compact objects (neutron stars and/or white dwarfs) or the accretion induced collapse of a white dwarf.  These evolutionary channels involve binary systems and are favoured by the high stellar density found in the cores of globular clusters. However,  the  implied   birth rates are relatively small thus  requiring  that FRB like the one in M81 must have active lifetimes larger than 10$^4$ yr  \citep{kre21,lu21}.  The lack of  strong bursting activity from \frb\ in the INTEGRAL data is consistent with this scenario  and supports  the view that not all FRBs  are related to young hyper-active magnetars.

\section{Conclusions} 
\label{sec:conclusions}

Our extensive search for hard X-ray magnetar-like bursts from \frb\   in the INTEGRAL/IBIS data obtained from November 2003  to  September 2021 did not reveal any significant event at the position of this repeating FRB. Being associated to a globular cluster in M81, \frb\ is by far  the closest among the well localized FRBs.   Given its distance of 3.6 Mpc, the average sensitivity of our search corresponds to  limits of   $\sim10^{43}$ erg on the isotropic energy  and   $\sim10^{45} \left ( \frac{10~ms}{\Delta t} \right )$ erg s$^{-1}$ on the 20-200 keV luminosity, where $\Delta$t is the burst duration. We can thus exclude the emission of giant flares during the time periods in which \frb\ was in the IBIS field of view ($\sim$18 Ms, in total). This supports the idea  that  \frb\ is not a young hyper-active magnetar, as also suggested by its lower luminosity and location in a globular cluster. 

Although  \frb\  might not be representative of the  bulk of the FRB  population, it is a very interesting target for multiwavelength campaigns.  Coordinated radio and  high-energy  observations of this source,  and other FRBs possibly associated to nearby galaxies (e.g.  FRB 20181030A,  \citealt{bha21b}),  can significantly  constrain models for the emission of radio bursts and possibly lead  to the discovery of the first  bursts from an extragalactic FRB. \\

 \bigskip
 
We thank A. Paizis for maintaining the INTEGRAL Archive at IASF-Milano,  A. Belfiore for help with the data analysis, and  the referee for useful comments.
This work has been  supported through the  ASI/INAF Agreement n. 2019-35-HH and PRIN-MIUR 2017 UnIAM (Unifying Isolated and Accreting Magnetars, PI S.~Mereghetti). 
Based on observations with INTEGRAL, an ESA project with instruments and science data centre funded by ESA member states (especially the PI countries: Denmark, France, Germany, Italy, Switzerland, Spain) and with the participation of the Russian Federeation and the USA.   ISGRI has been realized and maintained in flight by CEA-Saclay/Irfu with the support of CNES.



\bibliographystyle{aasjournal} 

%





\end{document}